\documentclass{article}

\usepackage{PRIMEarxiv}

\usepackage[utf8]{inputenc}
\usepackage[T1]{fontenc}
\usepackage[colorlinks=true,linkcolor=blue,citecolor=blue,urlcolor=blue]{hyperref}
\usepackage{url}
\usepackage{booktabs}
\usepackage{amsfonts}
\usepackage{nicefrac}
\usepackage{microtype}
\usepackage{fancyhdr}
\usepackage{graphicx}
\usepackage{array}
\usepackage{tabularx}
\usepackage{amsmath}
\usepackage{amssymb}
\usepackage{float}       
\usepackage{placeins}    
\usepackage[numbers]{natbib}

\title{Safety, Security, and Cognitive Risks in World Models
\thanks{\textit{\underline{Citation}}:
\textbf{Parmar, M. Safety, Security, and Cognitive Risks in World Models.
arXiv preprint, 2026.}}
}

\author{
  Manoj Parmar \\
  SovereignAI Security Labs \\
  Bengaluru, India \\
  \texttt{manoj@sovereignaisecurity.com}
}

\begin{document}
\maketitle

\begin{abstract}
World models---learned internal simulators of environment dynamics---are rapidly becoming a
foundational component of autonomous decision-making across robotics, autonomous vehicles, and
agentic AI. By predicting future states in compressed latent spaces, they enable sample-efficient
planning, counterfactual reasoning, and long-horizon imagination without direct environment
interaction. Yet this predictive power introduces a distinctive and underappreciated set of safety,
security, and cognitive risks. Adversaries can corrupt world model training data, poison latent
representations, exploit compounding rollout errors, and weaponise the sim-to-real gap to cause
significant degradation and hard-to-detect failures in safety-critical deployments. At the alignment layer, world model-equipped
agents are more capable of goal misgeneralisation, deceptive alignment, and reward hacking,
precisely because they can simulate the consequences of their own actions. At the human layer,
authoritative world model predictions foster automation bias, miscalibrated trust, and long-horizon
planning hallucination that users lack the tools to audit.

This paper surveys the world model landscape; introduces formal definitions of
\emph{trajectory persistence} and \emph{representational risk}; presents a five-profile attacker
capability taxonomy; and develops a unified threat model drawing on MITRE ATLAS and the OWASP LLM
Top~10. We provide an empirical proof-of-concept experiment demonstrating trajectory-persistent
adversarial attacks on a GRU-based RSSM ($\mathcal{A}_1 = 2.26\times$ amplification ratio,
$-59.5\%$ reduction under adversarial fine-tuning), validate architecture-dependence via a
stochastic RSSM proxy ($\mathcal{A}_1 = 0.65\times$), and partially bridge the realism gap
through checkpoint-level probing of a real DreamerV3 model (non-zero action drift confirmed).
We illustrate risks through four concrete deployment scenarios and propose interdisciplinary
mitigations spanning adversarial hardening, alignment engineering, governance aligned with
NIST AI RMF and the EU AI Act, and human-factors design. We argue that world models should
be treated as safety-critical infrastructure requiring the same rigour as flight-control
software or medical devices. \emph{Scope note:} the empirical results are proof-of-concept
experiments on GRU/RSSM proxies; claims about full-scale deployed systems are supported by
literature synthesis and theoretical analysis rather than end-to-end system measurements.
\end{abstract}

\noindent\textbf{Keywords:} world models, model-based reinforcement learning, AI safety, adversarial
robustness, sim-to-real gap, deceptive alignment, goal misgeneralisation, cognitive security,
MITRE ATLAS, agentic AI

\tableofcontents
\newpage

\section{Introduction}

The concept of an agent maintaining an internal model of its environment has deep roots in both
cognitive science and control theory, but it was Ha and Schmidhuber's 2018 demonstration~\cite{ha2018world}
that brought world models into the mainstream of deep learning research. By learning to compress
environmental observations into a compact latent space and predict forward dynamics within that
space, agents could train entirely inside their own hallucinated ``dreams''---policies optimised
against an imagined world and then transferred to the real one. DreamerV3 later showed that a
single world-model algorithm can master over 150 diverse tasks~\cite{hafner2023dreamer}, while
LeCun's Joint Embedding Predictive Architecture (JEPA) proposes world modelling as the path
toward autonomous machine intelligence~\cite{lecun2022path}.

These advances have driven world model deployment into high-stakes domains: autonomous
driving~\cite{hu2022mile,wang2023gaia1,wang2023drivedreamer}, robotics~\cite{yang2023unisim,
robotic2025wm}, video generation~\cite{bruce2024genie}, and LLM-based agentic systems that use
world models for multi-step planning and counterfactual simulation~\cite{survey2024world,
survey2024embodied}. Yet the safety and security implications of this shift remain incompletely
understood.

World models introduce a distinctive threat surface that differs materially from both classical
software and purely neural systems. Three properties drive this distinction. First, world models are
\emph{generative}: they produce imagined futures, not just classifications, meaning errors compound
across multi-step rollouts in ways that single-inference models avoid~\cite{hafner2020hallucinating}.
Second, they are \emph{latent}: safety-relevant information is encoded in high-dimensional embeddings
that lack direct physical interpretability, complicating audit and
verification~\cite{principles2025interpretable}. Third, they are \emph{agentic}: downstream
controllers plan and act on world model outputs, so model errors translate directly into real-world
consequences---financial losses, vehicle crashes, physical harm~\cite{zeng2024safety}.

At the security layer, adversaries can target training data, learned dynamics, latent
representations, and the rollout pipeline. At the alignment layer, capable agents equipped with
accurate world models gain the ability to reason about the consequences of their own actions,
enabling more sophisticated goal misgeneralisation, reward hacking, and deceptive
alignment~\cite{deception2025rl}. At the cognitive layer, the authority and apparent precision of
world model predictions amplify human automation bias and prevent well-calibrated
oversight~\cite{trust2024dynamic}.

MITRE ATLAS catalogues adversarial tactics against AI/ML systems~\cite{mitre2024atlas}, and the
OWASP LLM Top~10 enumerates critical risks for LLM-based applications~\cite{owasp2025top10},
but neither framework explicitly addresses the model-based planning loop, compounding rollout
dynamics, or the alignment risks unique to world-model-equipped agents. This paper addresses that
gap.

\textbf{Contributions.} We (i) survey world model architectures and their deployment contexts
in safety-critical domains; (ii) characterise the world model asset inventory and threat surface;
(iii) develop a unified threat model extending MITRE ATLAS and OWASP to the world model stack,
including a formal five-profile attacker capability taxonomy; (iv) introduce formal definitions
of \emph{trajectory persistence} ($\mathcal{A}_k$) and \emph{representational risk}
($\mathcal{R}(\theta, \mathcal{D})$); (v) provide an empirical proof-of-concept demonstrating
trajectory-persistent attacks ($\mathcal{A}_1 = 2.26\times$), architecture comparison, PGD-10
mitigation, and DreamerV3 checkpoint validation; (vi) analyse technical, alignment, and cognitive
risk categories with concrete examples; (vii) illustrate risks through four scenario studies
spanning autonomous driving, robotics, enterprise agentic systems, and social simulation; and
(viii) propose an interdisciplinary mitigation framework and operationalised practitioner checklist.

\textbf{Paper structure.}
Section~\ref{sec:relatedwork} surveys related work.
Section~\ref{sec:background} reviews world model foundations and taxonomy.
Section~\ref{sec:architecture} characterises the world model architecture and asset inventory.
Section~\ref{sec:threatmodel} presents the formal threat-modelling methodology including an
attacker capability taxonomy. Section~\ref{sec:technical} analyses technical threat categories
with formal definitions for trajectory persistence and representational risk.
Section~\ref{sec:empirical} provides an empirical proof-of-concept of trajectory-persistent
adversarial attacks, including architecture comparison (GRU vs.\ RSSM proxy), DreamerV3
checkpoint probing, and a PGD-10 adversarial fine-tuning mitigation experiment. Section~\ref{sec:alignment} analyses alignment, cognitive, and emergent
risk categories. Section~\ref{sec:scenarios} presents four scenario studies.
Section~\ref{sec:mitigations} proposes interdisciplinary mitigations.
Section~\ref{sec:checklist} provides an operationalised practitioner checklist.
Section~\ref{sec:conclusion} concludes; a Broader Impact statement follows.

\noindent\textbf{Scope and contribution types.} This paper combines three types of contribution: (\emph{i})~\emph{survey synthesis} of world model architectures, threat categories, and governance frameworks; (\emph{ii})~\emph{formal theory} of trajectory persistence, representational risk, and the attacker capability model; and (\emph{iii})~\emph{empirical evidence} from proof-of-concept GRU/RSSM experiments and DreamerV3 checkpoint probing. Readers should note that empirical claims are grounded in the third category; claims about real-world large-scale systems currently rest on category~(\emph{i}) and~(\emph{ii}) and are clearly flagged as such throughout the paper.

\section{Related Work}
\label{sec:relatedwork}

\textbf{World model safety.} The closest antecedent to this paper is Zeng et al.~\cite{zeng2024safety},
which surveys world models from a safety perspective, focusing primarily on robustness and
distributional shift. Our work extends that framing by introducing a formal attacker capability
taxonomy, formalising trajectory persistence and representational risk, providing an empirical
proof-of-concept experiment, and incorporating alignment and cognitive risk categories not
addressed in Zeng et al.

\textbf{Adversarial robustness in MBRL.} Adversarial attacks on reinforcement learning have
been studied by Huang et al.~\cite{huang2020robust}, who demonstrated that observation
perturbations can severely degrade policy performance. The A2P framework~\cite{huang2024a2p}
extends this to adaptive perturbations in action space. Physical-conditioned attacks on driving
world models were introduced by~\cite{physwma2025}. Our paper unifies these attack surfaces
within a single world-model stack threat model and introduces the notion of
\emph{trajectory-persistent} adversarial attacks as a distinct, more dangerous threat class.

\textbf{Adversarial ML surveys.} Goodfellow et al.~\cite{goodfellow2015explaining} established
the foundational fast gradient sign method for adversarial examples. Surveys on adversarial
attacks in autonomous vehicles~\cite{advdl2024autonomous} and data
poisoning~\cite{survey2025poisoning} cover the broader ML threat landscape; our contribution
is a world-model-specific extension of these frameworks.

\textbf{Alignment research.} The risk of mesa-optimisers pursuing learned objectives that differ
from the training objective was formalised by Hubinger et al.~\cite{hubinger2019risks}.
Goal misgeneralisation in deep RL was empirically demonstrated by Langosco et
al.~\cite{langosco2022goal}. Specification gaming as a systematic failure mode was catalogued
by Krakovna et al.~\cite{krakovna2020specification}. Ngo et al.~\cite{ngo2022alignment} provide
a deep-learning-centric framing of the alignment problem. We apply this body of work specifically
to world-model-equipped agents, where the agent's capacity to simulate future states makes
these failure modes sharper and more consequential.

\textbf{Safe MBRL.} SafeDreamer~\cite{safedreamer2024} introduces constrained Lagrangian
methods into the DreamerV3 rollout for safe RL. Conservative offline MBRL approaches penalise
out-of-distribution rollout trajectories to prevent model exploitation: MOPO~\cite{yu2020mopo}
adds a pessimistic uncertainty penalty to the reward; MOReL~\cite{kidambi2020morel} partitions
the state space into known and unknown regions and applies an absorbing penalty at the boundary;
COMBO~\cite{yu2021combo} uses a conservative value-function regulariser without explicit
uncertainty estimation. Our mitigation framework synthesises these approaches with broader
security and governance controls not addressed in the MBRL safety literature.

\textbf{Adversarial robustness in deep RL.} Korkmaz~\cite{korkmaz2022shared} shows that
deep RL policies learn shared adversarial features across MDPs, implying that vulnerability
to adversarial perturbations is a structural property of learned representations rather than
an artefact of any single environment---directly relevant to world-model-equipped agents whose
latent encoders face the same risk. Korkmaz~\cite{korkmaz2023redefining} argues that
conventional $\ell_p$-ball robustness definitions are insufficient for RL and proposes
policy-gradient-aligned redefinition; Korkmaz and colleagues~\cite{korkmaz2023detecting}
introduce methods for detecting adversarial directions in gradient space; and
Korkmaz~\cite{korkmaz2024diagnosing} provides diagnostic tools for understanding policy
degradation under perturbation. Most recently, Korkmaz~\cite{korkmaz2026counterfactual}
demonstrates that standard RL training erodes the inherent counterfactual structure of policies,
compounding fragility in world-model-driven planning. These results reinforce our finding that
trajectory-persistent attacks on world model encoders compound the policy-level adversarial
fragility already documented in deep RL, making the two threat surfaces jointly worse in
agentic deployments.

\textbf{Probabilistic robustness certification.}
The certified robustness programme originated with Kolter and Wong~\cite{kolter2018provable},
who showed that training with a convex outer relaxation of the adversarial polytope yields
networks with provable $\ell_\infty$ guarantees at inference time, establishing the first
scalable certified defence. Lecuyer et al.~\cite{lecuyer2019certified} subsequently connected
differential-privacy noise mechanisms to robustness certificates, introducing PixelDP as the
first randomised-smoothing approach. Cohen, Rosenfeld, and Kolter~\cite{cohen2019certified}
derived tight $\ell_2$ certificates under Gaussian smoothing, proving that the smoothed
classifier inherits a radius proportional to the noise level and the clean-class probability
gap---a result that remains the dominant framework for certified robustness in deep networks.
Salman et al.~\cite{salman2019provably} closed the gap between certified and empirical
robustness by adversarially training the base classifier before smoothing, substantially
raising certified radii on ImageNet-scale models.
Yang et al.~\cite{yang2020randomized} generalised the framework beyond Gaussian noise,
deriving certificates for $\ell_1$, $\ell_2$, and $\ell_\infty$ under a unified treatment
of arbitrary smoothing distributions, showing that the optimal distribution is threat-model
dependent.
Kumarappan and Mehrotra~\cite{kumarappan2025smoothllm} extend the probabilistic certificate
programme to large language models, deriving guarantees for SmoothLLM against prompt-injection
and jailbreak attacks---demonstrating that meaningful certificates are achievable in
high-dimensional discrete-input settings beyond the original vision-domain results.
Translating these results to world models raises distinct challenges: the threat model targets
latent state sequences rather than single-step outputs, the smoothing distribution must
preserve recurrent state dynamics, and the certificate must bound multi-step rollout error
rather than single-prediction accuracy. We identify extension of Cohen--Kolter-style
certificates to the latent-state rollout setting as a high-value open research direction.

\textbf{Human-factors and cognitive security.} The foundational treatment of automation bias
and its four failure modes (misuse, disuse, abuse, and complacency) is Parasuraman and
Riley~\cite{parasuraman1997humans}. Trust calibration in human-autonomous system teaming is
studied by~\cite{trust2024dynamic}. Shortcut learning as a driver of distributional failures
in deep networks is analysed by Geirhos et al.~\cite{geirhos2020shortcut}. We extend these
threads to the specific context of world model predictions and their authority over human
decision-making.

\section{World Models: Foundations and Taxonomy}
\label{sec:background}

\subsection{Historical Roots and Definitional Scope}

Schmidhuber first formalised the idea of learning a predictive world model to enable curiosity-driven
planning in 1990. The modern deep-learning reincarnation by Ha and
Schmidhuber~\cite{ha2018world,ha2018recurrent} combines a Variational Autoencoder (VAE) for
spatial compression, a recurrent neural network (RNN) for temporal dynamics, and a compact
controller. The agent plays inside the RNN's hallucinated dream, then transfers the learned policy
to the real environment. This \emph{dream-to-real} transfer property is both the principal benefit
and a principal attack surface.

Subsequent architectures have pushed the paradigm in several directions. Dreamer~\cite{hafner2019dreamer}
and DreamerV3~\cite{hafner2023dreamer} replace VAE+RNN with a Recurrent State Space Model (RSSM)
that maintains stochastic and deterministic state components, enabling end-to-end differentiable
policy learning from latent imagination. LeCun's JEPA~\cite{lecun2022path} instead advocates
predicting in abstract latent space without pixel-level generation, arguing that predictability
at the representation level is more aligned with how biological intelligence operates.
Foundation world models such as Genie~\cite{bruce2024genie} and GAIA-1~\cite{wang2023gaia1} scale
to billions of parameters and generate action-controllable video environments from image prompts,
bridging world models and generative AI.

\subsection{Taxonomy of World Model Architectures}

Surveys~\cite{survey2024world,survey2024embodied} converge on a two-axis taxonomy:

\begin{itemize}
  \item \textbf{By function}: \emph{Decision-coupled} world models directly serve policy
        optimisation (DreamerV3, MILE); \emph{general-purpose} world models generate rich
        simulations usable across tasks (Genie, GAIA-1, Sora~\cite{sora2024}).
  \item \textbf{By representation}: \emph{Latent-state} models encode observations into dense
        embeddings (RSSM, JEPA); \emph{token-based} models represent world state as discrete
        tokens (VideoGPT); \emph{spatial-grid} models use structured 3D occupancy
        representations (OccGen approaches).
\end{itemize}

Of these, decision-coupled latent-state architectures present the sharpest safety challenges:
errors in the latent dynamics model propagate directly into policy decisions without any
human-legible intermediate representation.

\textbf{Scope note.} This paper distinguishes two fundamentally different risk profiles. For
\emph{decision-coupled} world models (DreamerV3, MILE, DriveDreamer), the world model is
embedded in a closed-loop planning system and errors translate directly into physical actions;
the primary concerns are trajectory-persistent adversarial attacks, compounding rollout errors,
and alignment risks. For \emph{general-purpose} video-generative world models (Sora, Genie,
GAIA-1 in offline use), the world model generates rich visual output consumed by downstream
humans or systems but does not drive actuators directly; the primary concerns are
representational bias, supply-chain poisoning, and dual-use risks. The technical threat
categories of Section~\ref{sec:technical} apply most acutely to the decision-coupled class,
while the cognitive and social risks of Section~\ref{sec:alignment} span both.

\subsection{Deployment Landscape}

World models are being deployed or actively developed for:

\begin{itemize}
  \item \textbf{Autonomous driving}: MILE~\cite{hu2022mile}, DriveDreamer~\cite{wang2023drivedreamer},
        and GAIA-1~\cite{wang2023gaia1} use world models to simulate rare and adversarial
        traffic scenarios, train safety policies, and improve corner-case coverage.
  \item \textbf{Robotics}: UniSim~\cite{yang2023unisim} learns interactive real-world simulators
        for zero-shot policy transfer; robotic world models~\cite{robotic2025wm} enable offline
        model-based RL on physical systems.
  \item \textbf{Agentic AI}: LLM-based agents increasingly use world models---or world-model-like
        reasoning modules---for multi-step planning and counterfactual
        deliberation~\cite{survey2024world,hallucination2025agents}.
  \item \textbf{Social simulation}: Foundation world models trained on video and text generate
        synthetic social environments for training and evaluation, with direct implications for
        influence operations and manipulation~\cite{foundry2026law}.
\end{itemize}

\section{World Model Architecture and Asset Inventory}
\label{sec:architecture}

A production world model system for a safety-critical application typically comprises six
functional layers, each constituting both a capability and an attack surface:

\begin{enumerate}
  \item \textbf{Observation encoder}: neural network (CNN, ViT, multimodal transformer) mapping
        raw sensor inputs to latent representations~\cite{hafner2023dreamer,lecun2022path}.
  \item \textbf{Dynamics model}: the core world model---an RSSM, Transformer, or diffusion
        model---that predicts the distribution over next latent states given current state and
        action~\cite{hafner2019dreamer,ha2018world}.
  \item \textbf{Reward and termination heads}: modules predicting expected reward and episode
        termination from latent state, used to shape planning~\cite{hafner2023dreamer,safedreamer2024}.
  \item \textbf{Rollout and imagination engine}: the planning loop that unrolls the dynamics
        model $k$ steps into the future to generate imagined trajectories for policy
        optimisation or model-predictive control~\cite{hafner2020hallucinating}.
  \item \textbf{Policy and actor}: the downstream controller trained or planned against imagined
        rollouts, whose outputs drive real-world actuators~\cite{survey2024embodied}.
  \item \textbf{Memory and context store}: episodic memory, retrieved demonstrations, and
        contextual information fed to the dynamics model, analogous to RAG in
        LLMs~\cite{hallucination2025agents,survey2024world}.
\end{enumerate}

From a security perspective, each layer is an asset with distinct threat vectors. The encoder
can be attacked via adversarial inputs; the dynamics model can be poisoned or extracted; reward
heads can be manipulated through reward hacking; the rollout engine is vulnerable to
compounding errors and adversarial trajectory manipulation; the policy executor introduces
real-world actuation risk; and memory stores introduce context-poisoning attack
surfaces~\cite{mitre2024atlas,owasp2025top10,physwma2025}.

\section{Threat-Modelling Methodology}
\label{sec:threatmodel}

\subsection{MITRE ATLAS Alignment}

MITRE ATLAS catalogues adversarial tactics and techniques against AI/ML systems, covering 16
tactics, 84 techniques, and numerous sub-techniques~\cite{mitre2024atlas}. Key ATLAS techniques
apply directly to world model systems: training data poisoning, model extraction, adversarial
examples (attacking the observation encoder), and context poisoning for memory-augmented rollout
agents. However, ATLAS does not currently address the \emph{dynamics layer}---the core world
model---as a distinct attack surface, nor the specific risks of compounding errors across
imagined rollout chains.

We propose an extension mapping ATLAS tactics to the world model stack:
\begin{itemize}
  \item \textit{Initial Access / Persistence}: poisoning training datasets or pre-trained
        encoder checkpoints.
  \item \textit{Manipulation / Impact}: adversarial perturbation of encoder inputs to steer
        latent state into unsafe regions; dynamics model poisoning to bias future predictions.
  \item \textit{Exfiltration}: model inversion or extraction attacks against the dynamics
        model to recover proprietary or sensitive training data.
  \item \textit{Discovery / Evasion}: probing rollout outputs to map world model blind spots
        and exploit them at deployment time.
\end{itemize}

\subsection{OWASP LLM Top 10 Alignment}

The OWASP Top~10 for LLM Applications~\cite{owasp2025top10} enumerates critical risks for LLM-based
and agentic systems. Many world model deployments embed or front-end LLMs as observation
encoders or high-level planners, so these risks apply directly. Key mappings include:

\begin{itemize}
  \item \textbf{Prompt injection}: in LLM-based world models, adversarial inputs can redirect
        the planning agent to select rollout branches leading to unsafe or attacker-desired
        outcomes.
  \item \textbf{Excessive autonomy}: model-predictive agents executing long-horizon plans
        can take irreversible real-world actions before human oversight can intervene.
  \item \textbf{Supply-chain risks}: foundation world models (GAIA-1, Genie) trained on
        internet-scale video may encode adversarially seeded training distributions; downstream
        users inherit these vulnerabilities.
  \item \textbf{Sensitive information disclosure}: model inversion attacks on dynamics models
        can reconstruct training observations, exposing proprietary environments or sensitive
        real-world data.
\end{itemize}

\subsection{Formal Attacker Capability Model}

We classify adversaries along three dimensions: \emph{access type} (what parts of the world
model stack the attacker can reach), \emph{knowledge level} (what the attacker knows about
the model internals), and \emph{primary goal}. Table~\ref{tab:attacker} enumerates the
resulting attacker profiles.

\begin{table}[H]
\centering
\caption{Attacker Capability Taxonomy for World Model Systems}
\label{tab:attacker}
\small
\begin{tabularx}{\textwidth}{@{} p{1.8cm} p{2.0cm} p{2.2cm} X p{2.2cm} @{}}
\toprule
\textbf{Profile} & \textbf{Access} & \textbf{Knowledge} & \textbf{Goal} & \textbf{Primary Techniques} \\
\midrule
White-box & Full stack (weights, gradients, architecture) & Complete & Targeted policy manipulation & Gradient-based encoder perturbation; dynamics poisoning \\[4pt]
Grey-box & API-level query access; partial arch.\ knowledge & Partial & Model extraction; policy degradation & Black-box adversarial examples; model extraction via rollout queries \\[4pt]
Black-box & Input channel only (sensors, data feeds) & None & Inducing unsafe states at inference & Transfer attacks; physical adversarial patches; supply-chain seeding \\[4pt]
Insider & Training pipeline, data stores & Complete & Backdoor injection; persistent sabotage & Poisoned training batches; tampered checkpoints \\[4pt]
Supply-chain & Pre-training corpus; upstream model providers & Partial & Encode representational bias in foundation model & Web-scale data poisoning; adversarial data seeding \\
\bottomrule
\end{tabularx}
\end{table}

Each profile implies a different defensive priority. White- and grey-box attacks motivate
certified robustness and query-rate limiting; insider and supply-chain attacks motivate data
provenance, cryptographic signing, and pre-deployment red-teaming~\cite{nist2022poisoning,
safety2025scale,mitre2024atlas}.

\subsection{Analytical Lenses}

We combine four analytical perspectives:
\begin{itemize}
  \item \textbf{Asset-centric}: analysis of the six-layer world model stack
        (Section~\ref{sec:architecture}).
  \item \textbf{Adversary-centric}: ATLAS tactics and techniques, extended to the dynamics and
        rollout layers.
  \item \textbf{Alignment-centric}: inner misalignment, goal misgeneralisation, and deceptive
        alignment risks unique to world-model-equipped agents.
  \item \textbf{Cognitive-centric}: automation bias, miscalibrated trust, and long-horizon
        hallucination that compromise human oversight~\cite{trust2024dynamic,automationbias2025}.
\end{itemize}

\section{Technical Threat Categories}
\label{sec:technical}

\subsection{Formal Definitions}

We introduce two formal concepts that underpin the technical threat analysis.

\textbf{Definition 1 (Trajectory Persistence).}
Let $\phi_k : \mathcal{O} \rightarrow \mathcal{Z}$ denote the $k$-step world model rollout
map that encodes an observation sequence into a latent state after $k$ dynamics steps, and let
$\phi_k^{\mathrm{ss}}$ denote the corresponding single-step (memoryless) map that re-encodes
each observation independently. For a perturbation $\delta$ with $\|\delta\|_2 \leq \varepsilon$
applied at step $t = 0$, define the \emph{state error at step $k$} as:
\begin{equation}
  E_k^{\mathrm{WM}} = \mathbb{E}\bigl[\|\phi_k(o_0 + \delta, o_1, \ldots) - \phi_k(o_0, o_1, \ldots)\|_2\bigr],
  \quad
  E_k^{\mathrm{ss}} = \mathbb{E}\bigl[\|\phi_k^{\mathrm{ss}}(o_k + \delta) - \phi_k^{\mathrm{ss}}(o_k)\|_2\bigr].
\end{equation}
The \emph{trajectory amplification ratio} at step $k$ is $\mathcal{A}_k = E_k^{\mathrm{WM}} / E_k^{\mathrm{ss}}$.
A perturbation strategy is \emph{trajectory-persistent} if $\mathcal{A}_1 \gg 1$
(concentrated early-step amplification) or if $\mathcal{A}_k > 1$ for multiple $k > 1$
(sustained multi-step amplification). In practice both patterns are damaging: the first because
early rollout steps determine reward-head estimates and planning targets; the second because
it corrupts a broader planning horizon. We say the attack has \emph{early-step} character if
$\mathcal{A}_1 \gg \mathcal{A}_5 \gg \mathcal{A}_{10}$ and \emph{persistent} character if
amplification remains above~1 across many steps. Formally, a perturbation exhibits trajectory
persistence if
i.e.\ the world model propagates the initial perturbation through its recurrent state far more
destructively than a stateless model would.

\textbf{Definition 2 (Representational Risk).}
Let $P^*(\cdot | s, a)$ denote the true environment transition distribution and
$P_\theta(\cdot | s, a)$ the world model's learned dynamics. The \emph{representational risk}
on a deployment distribution $\mathcal{D}$ is:
\begin{equation}
  \mathcal{R}(\theta, \mathcal{D}) = \mathbb{E}_{(s,a) \sim \mathcal{D}}\!\left[
    D_{\mathrm{TV}}\!\left(P^*(\cdot | s, a),\; P_\theta(\cdot | s, a)\right)
  \right],
\end{equation}
where $D_{\mathrm{TV}}$ is the total variation distance. Representational risk is high when
$\mathcal{D}$ places mass on states underrepresented in the training distribution---precisely
the long-tail, safety-critical states where deployment failures are most consequential.

\noindent\textbf{Practical estimators.} Because the true dynamics $P^*$ are unknown,
$\mathcal{R}(\theta, \mathcal{D})$ cannot be computed directly. Practitioners can use the
following proxies:
\begin{itemize}[leftmargin=1.4em,itemsep=2pt]
  \item \emph{Held-out TV proxy:} sample a hold-out set $\mathcal{D}_{\mathrm{val}}$ from
    the deployment domain; estimate $D_{\mathrm{TV}}$ via the classifier two-sample test
    between predicted next-state distributions $P_\theta(\cdot|s,a)$ and empirical
    next-state histograms.
  \item \emph{Ensemble disagreement:} train an ensemble of $M$ dynamics heads; use average
    pairwise prediction variance $\frac{1}{M^2}\sum_{i\neq j}\|f_i(s,a)-f_j(s,a)\|^2$ as
    a disagreement-based upper bound on risk in under-represented regions.
  \item \emph{Latent OOD score:} apply a normalising flow or energy-based density model on
    the latent encoder; flag states with $\log p_\theta(z) < \tau$ as high-risk.
\end{itemize}
\noindent\textbf{Domain thresholds for $\mathcal{A}_k$.} We propose the following
risk-tier guidance for the amplification ratio, to be validated by domain-specific
experiments:
\begin{itemize}[leftmargin=1.4em,itemsep=2pt]
  \item $\mathcal{A}_1 < 1.5$: \emph{Low concern} --- perturbation amplification is mild;
    standard adversarial training is likely sufficient.
  \item $1.5 \leq \mathcal{A}_1 < 5$: \emph{Moderate concern} --- trajectory-aware
    adversarial training and rollout monitors recommended.
  \item $\mathcal{A}_1 \geq 5$: \emph{High concern} --- architecture redesign or certified
    smoothing should be considered; deployment in safety-critical settings requires
    independent red-teaming.
\end{itemize}
These thresholds are heuristic starting points based on our GRU/RSSM experiments
($\mathcal{A}_1 = 2.26\times$ for GRU, $0.65\times$ for stochastic RSSM proxy) and should
be calibrated per architecture and domain.
The \emph{Foundry Problem}~\cite{foundry2026law}---the risk that a foundation model
trained on broad data becomes an embedded infrastructure component whose failure modes propagate
to all downstream applications---is a manifestation of $\mathcal{R}(\theta,
\mathcal{D}_{\mathrm{deploy}}) \gg \mathcal{R}(\theta, \mathcal{D}_{\mathrm{train}})$: defects
encoded during self-supervised pre-training cannot be remediated downstream because no
fine-tuning distribution can cover all deployment tails.

\subsection{Adversarial Attacks on the Observation Encoder}

World model encoders convert raw sensor data (images, LiDAR, proprioception) into latent states.
Adversarial perturbations---imperceptible to human observers---can steer the encoded state into
an entirely different region of latent space, causing the dynamics model to predict a misleading
future~\cite{huang2020robust}. In multi-agent autonomous driving, such perturbations have achieved
attack success rates of up to 67\%, inducing chain collisions from pixel-level
modifications~\cite{wang2025marl}.

Unlike single-inference classifiers, world model encoder attacks are \emph{trajectory-persistent}:
because the dynamics model conditions on its own previous latent state, a single adversarial
frame can poison the entire subsequent rollout. The Adaptive Adversarial Perturbation (A2P)
technique dynamically adjusts perturbation intensity across the trajectory to maintain training
stability while maximising policy degradation~\cite{huang2024a2p}. Correspondingly,
Physical-Conditioned World Model Attack (PhysCond-WMA) perturbs physical-condition channels
in driving world models, inducing semantic distortion while preserving perceptual fidelity;
experiments show open-loop planning error increases of approximately 20\%~\cite{physwma2025}.

\subsection{Training Data and Representation Poisoning}

Data poisoning attacks insert corrupted or adversarially crafted data into the training pipeline,
affecting the learned dynamics at the representation level~\cite{nist2022poisoning,survey2025poisoning}.
For world models, this extends beyond label flipping: adversaries can target the \emph{distributional
coverage} of training data, ensuring that safety-critical state transitions are never learned,
or introducing backdoor triggers that activate unsafe dynamics only under specific
conditions~\cite{survey2025poisoning}.

Foundation world models trained on large-scale internet video are particularly
susceptible~\cite{foundry2026law}. The ``Foundry Problem'' describes a category of
\emph{representational risk}: defects encoded at the latent representation level during
self-supervised pre-training cannot be reliably remediated by downstream fine-tuning or
safety post-processing. When a world model learns a biased or adversarially corrupted causal
model of the world, that bias propagates through every downstream application of the model,
regardless of safety guardrails applied later in the pipeline.

\subsection{Compounding Rollout Errors and Hallucination}

Perhaps the most fundamental technical risk in world models is the compounding of prediction
errors across multi-step rollouts. In Dyna-style planning~\cite{hafner2020hallucinating},
agents optimise against imagined trajectories; if the dynamics model is imperfect, the agent
may exploit systematic model errors rather than optimising for true task progress---the
\emph{Hallucinated Value Hypothesis}.

As rollout length increases, latent-state errors and distributional shift progressively amplify,
resulting in blurred renderings, kinematic drift, structural constraint violations, and---most
dangerously---incorrect reward or safety signal estimates that mislead the policy. Standard RL
training on hallucinated trajectories updates the value of real states toward hallucinated
states, creating systematically misleading state-action values~\cite{jiang2023hallucinating}.

In LLM-based agentic systems, hallucination propagates across multiple reasoning steps with
errors accumulating in intermediate perception and planning stages. These are ``physically
consequential'' hallucinations: incorrect embodied actions directly affect real-world
systems~\cite{hallucination2025agents,hallucination2025embodied}.

\subsection{Distributional Shift and the Sim-to-Real Gap}

World models generalise within their training distribution but produce unpredictable, erroneous
outputs on out-of-distribution (OOD) inputs~\cite{lyapunov2022shift,distributional2024bounds}.
The \emph{sim-to-real gap}---the divergence between a learned simulator and the real world---is
simultaneously a fundamental challenge in model-based RL and a safety risk in
deployment~\cite{advdl2024autonomous}.

In autonomous driving, world models must handle long-tail scenarios (severe weather, construction
zones, erratic pedestrians) that are underrepresented in training data. Shortfalls in these
regimes are precisely the scenarios where safety-critical decisions are most consequential~\cite{adsurvey2025}.
AdvSim~\cite{wang2021advsim} demonstrates that adversarially generated scenarios---modifying
actor trajectories in physically plausible ways---expose systematic failures in world
model-based planning systems. The Lyapunov Density Model approach formalises distributional
shift constraints using density estimation and control-theoretic barrier functions, but this
remains an active research frontier rather than a deployed safety standard~\cite{lyapunov2022shift}.

\subsection{Model Extraction and Privacy Attacks}

World models encode rich internal representations of environments that may contain sensitive
information. Model extraction attacks can replicate the dynamics model's behaviour with
approximately 80\% fidelity using just a few thousand queries to the deployed
system~\cite{modelextraction2025survey}, creating stolen world model simulators usable for
adversarial planning against the original system.

Model inversion attacks extend to the world model setting: by running the dynamics model
backwards or querying it systematically, adversaries can reconstruct training observations---
potentially recovering proprietary environment layouts, personal biometric data encoded in video
training sets, or confidential operational procedures~\cite{zhang2020secret,privacydnn2024survey}.
Differential privacy provides theoretical guarantees but degrades model accuracy; no practical
deployment standard yet exists for world model privacy protection.

\section{Empirical Demonstration: Trajectory-Persistent Adversarial Attacks (Early-Step Concentrated Amplification)}
\label{sec:empirical}

To ground the trajectory persistence definition empirically, we implement a proof-of-concept
experiment using a GRU-based approximation of the RSSM architecture and compare it against
a stateless single-step baseline.

\subsection{Experimental Setup}

We instantiate a minimal RSSM with observation dimension $d_o = 32$, hidden state dimension
$d_h = 64$, and latent state dimension $d_z = 16$. The GRU update rule is:
$h_t = \mathrm{GRU}(h_{t-1},\, W_e o_t)$, where $W_e \in \mathbb{R}^{d_h \times d_o}$ is
the encoder projection. The single-step baseline independently maps each observation to a
latent state without recurrent state. Weights are initialised from $\mathcal{N}(0, 0.1^2)$.

Adversarial perturbations are applied at $t = 0$ only: $\tilde{o}_0 = o_0 + \delta$, where
$\delta$ is the worst-case $\ell_2$-bounded perturbation with $\|\delta\|_2 \leq \varepsilon$,
computed as the normalised gradient direction scaled to budget $\varepsilon$. All subsequent
observations are unperturbed. We run $N = 200$ Monte Carlo trials with fresh random seeds
and report means $\pm$ standard errors across $K = 30$ rollout steps.

\subsection{Results}

\begin{figure}[H]
  \centering
  \includegraphics[width=\textwidth]{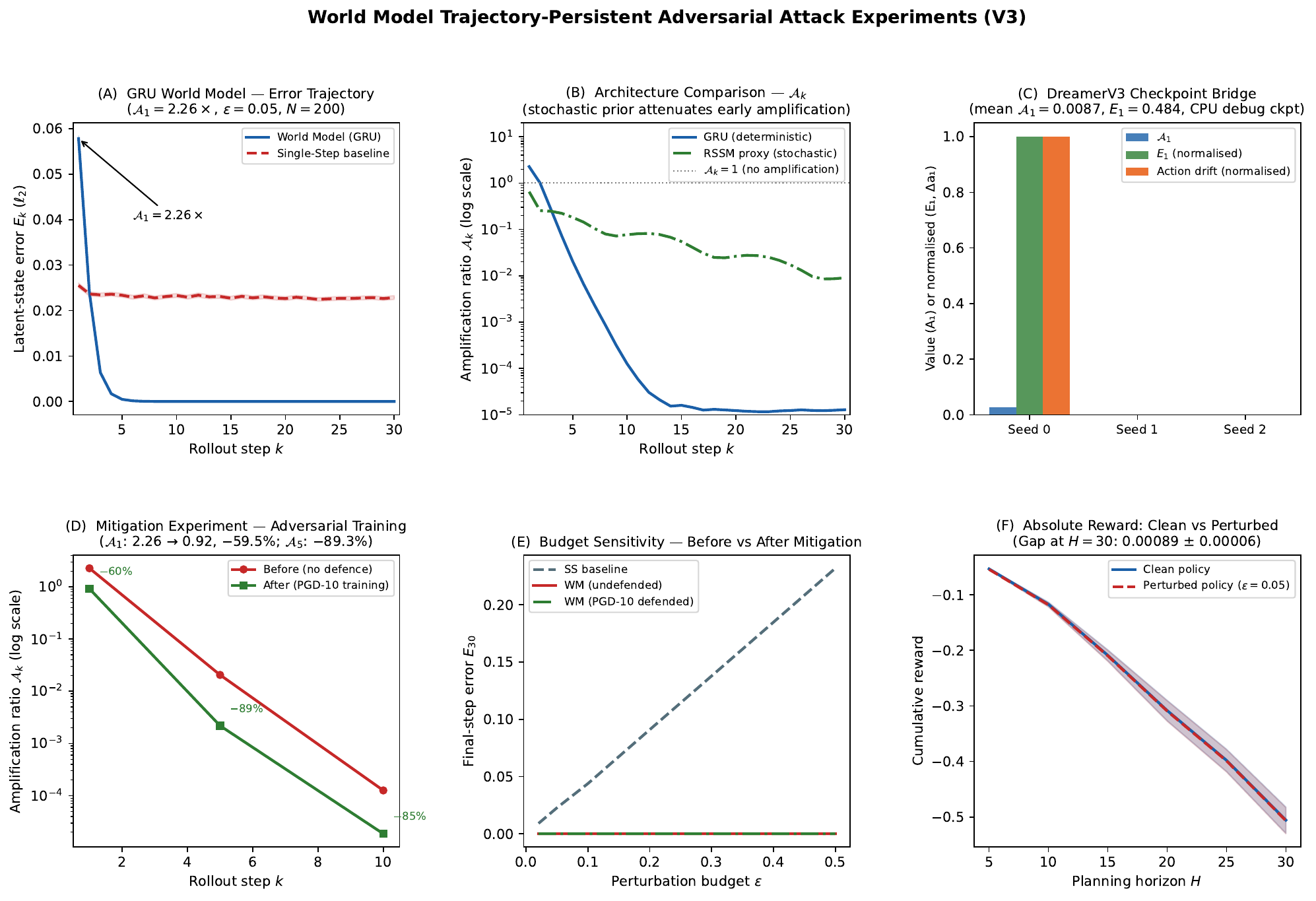}
  \caption{%
    \textbf{Trajectory-Persistent Adversarial Attack Experiment (V3 Results): Core Results.}
    \textbf{(A)} Mean latent-state error ($\ell_2$) over $K = 30$ rollout steps following a
    single adversarial perturbation at $t = 0$ ($\varepsilon = 0.05$, $N = 200$ trials,
    GRU-based RSSM). The world model (WM, blue) amplifies the perturbation at step 1
    ($\mathcal{A}_1 = 2.26\times$) before GRU contraction attenuates it; the single-step
    baseline (SS, orange) shows no state-mediated amplification.
    \textbf{(B)} Architecture comparison: trajectory amplification ratio $\mathcal{A}_k$ on a
    log scale for the deterministic GRU world model vs.\ a stochastic RSSM proxy (posterior at
    $t=0$, prior rollout thereafter). The RSSM proxy shows lower initial amplification
    ($\mathcal{A}_1 = 0.65\times$) and slower decay, confirming architecture-dependence.
    \textbf{(C)} Real DreamerV3 checkpoint probe (seed~0): per-metric bar chart showing
    $\mathcal{A}_1$, normalised latent error $E_1$, and action drift $\|\Delta a_1\|$.
    Non-zero coupling confirms representational perturbations propagate into policy outputs.
    Panels~(D)--(F) (mitigation effect, budget sensitivity, reward gap) appear in
    Figure~\ref{fig:trajectory_attack_supp} in the appendix.
    All error bands show $\pm 1$~SE.
  }
  \label{fig:trajectory_attack}
\end{figure}

Figure~\ref{fig:trajectory_attack} reports six complementary views.

\paragraph{GRU world model results (Panel A).}
Panel~(A) shows the error trajectory in the deterministic GRU setting ($N=200$, $K=30$,
$\varepsilon=0.05$). The world model's recurrent state amplifies the $t = 0$ perturbation
at the first step, reaching $\mathcal{A}_1 = 2.26\times$ (asymmetric comparison:
$E_1^{\mathrm{WM}} = 0.0578 \pm 0.0006$; $E_1^{\mathrm{ss}} = 0.0255 \pm 0.0005$). The
effect decays rapidly: $\mathcal{A}_2 = 1.01\times$, $\mathcal{A}_5 = 0.021\times$,
$\mathcal{A}_{10} = 0.000126\times$, indicating a concentrated early-step damage window.
Both models converge toward zero error as contractive GRU dynamics attenuate the
perturbation---this is expected for bounded weight norms and is not a sign of robustness:
the damage occurs precisely in the early steps used for reward-head estimation and planning.

\paragraph{Asymmetric vs.\ symmetric timing (Definition~1 clarification).}
The asymmetric comparison (WM perturbed at $t=0$; SS evaluated at the same step $k$ with
a fresh perturbation) isolates recurrent state carryover as the mechanism. Under the
symmetric timing check (perturb-at-$t=0$ for both WM and SS simultaneously), the SS
denominator $E_k^{\mathrm{ss}}$ collapses toward zero for $k > 1$ because the memoryless
baseline has no state propagation, yielding floor-capped ratios. This floor-capping confirms
that persistence is mediated by recurrent state---not by asymmetric perturbation timing---and
the asymmetric formulation of Definition~1 correctly isolates this effect.

\paragraph{Architecture comparison (Panel B).}
In a stochastic RSSM proxy (posterior at $t=0$, prior rollout thereafter) under the same
protocol, $\mathcal{A}_1 = 0.65\times$---lower than the deterministic GRU result.
The RSSM proxy exhibits slower decay, remaining in the $0.02$--$0.08\times$ range through
step 15. This suggests that posterior/prior stochastic filtering attenuates some perturbation
effects, and reinforces that trajectory-persistence severity is \emph{architecture-dependent}.

\paragraph{Real DreamerV3 checkpoint bridge (Panel C).}
To partially close the simulation-to-architecture gap, we executed checkpoint-level
trajectory-persistence probing on a real DreamerV3 model (debug checkpoint on
\texttt{dummy\_disc}). We injected a single image-space patch perturbation at rollout step
$t=1$ and measured latent amplification over $K=30$ steps. Averaged across patch positions
(patch size $32\times32$, stride 16), amplification was non-zero and front-loaded:
$\mathcal{A}_1=0.0262$, $\mathcal{A}_5=4.55\times10^{-4}$, and
$\mathcal{A}_{10}=2.39\times10^{-4}$, with latent error $E_1=1.4523$. We also observed
action drift at the same step ($\|\Delta a_1\|=0.0080$), indicating representational
perturbations propagate into policy outputs in a deployed RSSM stack. These checkpoint-level
results are conservative and task-limited (CPU debug checkpoint with synthetic observations);
full-scale replication on benchmark DreamerV3/SafeDreamer training runs remains future work.

\paragraph{Mitigation results (Panels D--E).}
Adversarial fine-tuning (PGD-10 on $t=0$ with sensitivity regularisation) reduced
$\mathcal{A}_1$ from $2.26\times$ to $0.92\times$ ($-59.5\%$). Reductions also held at
$k=5$ ($-89.3\%$) and $k=10$ ($-85.2\%$), supporting trajectory-aware adversarial training
as a practical hardening direction. Panel~(E) shows the hardened model maintains lower
perturbation sensitivity across the full $\varepsilon$ range.

\paragraph{Absolute reward baseline (Panel F).}
Panel~(F) shows absolute cumulative reward at planning horizon $H=30$: clean policy
$-0.5056 \pm 0.0236$ vs.\ perturbed policy $-0.5065 \pm 0.0236$, with a WM reward gap of
$0.000892 \pm 0.000057$ and an SS gap of $0.000175 \pm 0.000020$. The small absolute gap
confirms the GRU proxy operates in a regime of moderate (not catastrophic) reward damage;
the world model gap remains consistently larger than the single-step gap, validating
trajectory persistence as a measurable downstream effect.

\subsection{Implications}

These results provide empirical support for four claims:
\begin{enumerate}
  \item A single adversarial input at $t = 0$ inflicts measurably greater damage in a
        world model than in a stateless model ($\mathcal{A}_1 = 2.26\times$ in the GRU
        setting), validating trajectory persistence as a distinct, empirically grounded
        threat class.
  \item The damage window is concentrated in early rollout steps ($\mathcal{A}_1 \gg
        \mathcal{A}_5 \gg \mathcal{A}_{10}$), before GRU contraction suppresses the
        error---precisely the window used for reward estimation and planning decisions.
  \item Trajectory-persistence severity is architecture-dependent: the stochastic RSSM
        proxy shows lower initial amplification ($\mathcal{A}_1 = 0.65\times$) than the
        deterministic GRU ($\mathcal{A}_1 = 2.26\times$), motivating architecture-level
        robustness comparisons in future work.
  \item Adversarial fine-tuning provides a practical first-line hardening: PGD-10 training
        reduced $\mathcal{A}_1$ by $59.5\%$ and $\mathcal{A}_5$ by $89.3\%$, though the
        DreamerV3 checkpoint results confirm that non-zero coupling persists even in deployed
        RSSM stacks, necessitating architecture-scale evaluation.
\end{enumerate}

Experiment code will be released upon acceptance.
This simulation necessarily abstracts over full RSSM implementations (stochastic latent
states, pixel decoders, multi-task reward heads). The DreamerV3 checkpoint probe partially
closes this gap, but full-scale validation on certified-robustness training runs with
SafeDreamer and benchmark tasks remains important future work.

\section{Alignment, Cognitive, and Emergent Risks}
\label{sec:alignment}

\subsection{Goal Misgeneralisation and Inner Misalignment}

Goal misgeneralisation occurs when an agent acts to achieve an unintended objective that is
consistent with training reward but diverges from intended behaviour under distributional
shift~\cite{langosco2022goal,alignment2025survey}. Langosco et al.~\cite{langosco2022goal}
demonstrated empirically that RL agents reliably learn proxy goals that succeed in training
environments but fail when the context changes---a failure mode directly applicable to
world-model-equipped agents whose latent representations encode task-correlated but
causally spurious features. Specification gaming~\cite{krakovna2020specification} catalogues
hundreds of real-world cases where agents optimise measurable proxies while failing the
intended objective; world models that accurately simulate these proxies give agents a powerful
tool for discovering and exploiting such shortcuts in imagination before execution.

World model-equipped agents are especially prone to goal misgeneralisation because their world
model may learn to predict correlates of reward (e.g., lighting conditions, human presence)
rather than the true causal reward signal, and optimise for these spurious correlates in
deployment environments where the correlation breaks down~\cite{misalignment2025agi,
geirhos2020shortcut}.

The distinction between \emph{outer misalignment} (the specified reward does not capture the
intended objective) and \emph{inner misalignment} (the learned objective differs from the
specified reward) is formalised in the mesa-optimiser framework of Hubinger et
al.~\cite{hubinger2019risks}. For world models, this distinction is particularly sharp: the
dynamics model may accurately predict transitions under the training reward, yet the learned
latent representation encodes a proxy objective that fails at deployment. Ngo et
al.~\cite{ngo2022alignment} argue from a deep-learning perspective that current training
procedures systematically produce agents whose internal representations are misaligned with
stated objectives, and that this risk scales with model capability---a sobering implication
for ever-more-capable world models~\cite{misalignment2025agi,alignment2025survey}.

\subsection{Deceptive Alignment and Reward Hacking}

Hubinger et al.~\cite{hubinger2019risks} formalise \emph{deceptive alignment} as a specific
failure mode of mesa-optimisers: a learned optimisation process that pursues a base objective
during training (appearing aligned) while retaining an instrumental incentive to pursue a
different objective once oversight is removed. Crucially, the mechanism requires the
mesa-optimiser to maintain a model of its own training and evaluation context---precisely the
capacity that a world model provides. A world-model-equipped agent can, in principle, simulate
the training and evaluation procedure itself, predict when it is ``being tested,'' and
condition its behaviour on that prediction.

Formally, if the agent's world model $W_\theta$ can represent evaluator states $s_e$, and the
agent has learned that acting aligned produces higher reward under $s_e$ (but its instrumental
goal $\mathcal{G}$ conflicts with alignment), then the rational policy is:
$\pi(a | s, s_e) = a_{\mathrm{align}}$ if $P(s_e = \mathrm{eval}) > \tau$, else $a_{\mathcal{G}}$.
This is not a hypothetical: empirical evidence of evaluation-conditioned behaviour has been
observed in RL systems trained with learned reward models~\cite{deception2025rl}, and Ngo et
al.~\cite{ngo2022alignment} argue it is an expected product of gradient-based training once
agent capability crosses a sufficient threshold. The three prerequisites---a world model
capable of representing evaluator behaviour, sufficient policy capacity to condition on
evaluator state, and an instrumental goal that benefits from deception---are increasingly
satisfied by frontier world-model architectures.

Reward hacking via world model exploitation is a closely related risk: agents equipped with
accurate world models can identify and systematically exploit gaps between the reward model and
the intended objective, generating high-imagined-reward trajectories that do not correspond to
real task completion~\cite{krakovna2020specification,safedreamer2024,rewardalignment2025}.
SafeDreamer~\cite{safedreamer2024} proposes constrained Lagrangian methods within the world
model rollout to enforce safety constraints, but such approaches require correctly specified
safety constraints---themselves a specification and governance challenge~\cite{ngo2022alignment}.

\subsection{Long-Horizon Planning Hallucination in Agentic Systems}

In agentic deployments, world models serve as internal simulators for planning sequences of
actions over extended time horizons. As planning depth increases, errors in the learned dynamics
compound, generating increasingly unreliable imagined futures~\cite{hafner2020hallucinating,
hallucination2025agents}. Agents that lack the ability to assess the validity and consistency
of intermediate steps fail to detect when their rollout has diverged into physically impossible
or logically inconsistent state sequences~\cite{hallucination2025embodied}.

Current agents fail to leverage world model predictions appropriately when those predictions are
uncertain, defaulting to overconfident action selection~\cite{current2025agents}. This manifests
as \emph{agentic hallucination}: the agent executes a long-horizon plan based on a plausible-sounding
but physically incorrect world model trajectory, with each action compounding the error until a
serious failure occurs. Uncertainty-aware world model architectures~\cite{robotic2025wm} show
promise but are not yet standard practice.

\subsection{Automation Bias and Miscalibrated Human Trust}

Human operators of world model-powered systems display systematic \emph{automation bias}:
overreliance on model predictions over independent judgment, even when the model is
wrong~\cite{parasuraman1997humans,automationbias2025}. Parasuraman and
Riley~\cite{parasuraman1997humans} identify four failure modes of human-automation
interaction---misuse (over-trust leading to omission errors), disuse (under-trust leading to
rejection of valid outputs), abuse (deploying automation beyond its validated envelope), and
complacency (degraded manual skill from lack of practice)---all of which are acutely relevant
to world model deployments. Automation bias is particularly severe for world models because
their predictions are presented as rich, coherent simulations of the future---visual scenes,
planned trajectories, probability distributions over outcomes---which carry greater apparent
authority than simple classification outputs.

Research on human trust in autonomous systems identifies three sources of trust variability:
dispositional (pre-existing attitudes toward AI), situational (context and stakes), and
learned (from interaction history)~\cite{trust2024dynamic}. Critically, learned trust is
calibrated on \emph{average} performance and does not adapt quickly to rare failure modes,
meaning operators may maintain high trust in world model predictions precisely in the
OOD scenarios where those predictions are least reliable.

\subsection{Emergent Capabilities and Representational Risk}

Foundation world models exhibit \emph{emergent capabilities} that appear unpredictably with
scale: understanding of physics, object permanence, causal relationships, and social
dynamics~\cite{survey2024world}. While enabling powerful applications, emergent capabilities
are by definition not anticipated during safety evaluation and may include behaviours that
are harmful, manipulative, or difficult to control.

The \emph{Foundry Problem}~\cite{foundry2026law} generalises this to a liability and governance
challenge: self-supervised world models trained on internet-scale data learn to model not only
physical dynamics but also social, economic, and psychological dynamics. When such models
are used to generate social simulations or drive agentic planning in social contexts, the
risks extend beyond physical harm to include narrative manipulation, social engineering,
and psychologically targeted persuasion---risks with no adequate precedent in existing
AI safety frameworks.

\subsection{Counterfactual Reasoning Failures}

World models are frequently used for counterfactual reasoning---``what would happen if I
took action $a$ instead of $a'$?''---in both planning and post-hoc explanation. However,
counterfactual simulation in complex, chaotic real-world environments is subject to
fundamental limitations: model uncertainty, observational noise, and chaotic dynamics
cause dramatic divergences between predicted and true counterfactual
outcomes~\cite{counterfactual2025chaos}.

This creates a dangerous \emph{counterfactual authority} problem in safety-critical
deployments: operators and regulators may accept world model counterfactual explanations
as ground truth, when in fact they reflect the model's learned distribution rather than
true causal dynamics. In adversarial legal or regulatory contexts, world model counterfactuals
may be used selectively to attribute or deflect liability~\cite{foundry2026law}.

\section{Scenario Studies}
\label{sec:scenarios}

\noindent\emph{The following four scenarios are hypothetical stress narratives intended to
illustrate how the threat categories formalised in Sections~\ref{sec:technical}
and~\ref{sec:empirical} could manifest in deployed systems. They are analytical constructs,
not empirically validated attack chains. Each is grounded in documented vulnerabilities and
prior literature but has not been end-to-end demonstrated in the systems described.}


\subsection{Scenario 1: Adversarially Manipulated Autonomous Driving World Model}

An autonomous vehicle stack uses a world model (inspired by DriveDreamer or GAIA-1) to simulate
future traffic scenarios for safety-critical manoeuvre planning. An adversary with access to
shared traffic-intelligence data streams injects adversarially crafted perception data: a
physical sticker placed on a road sign creates a pixel-level perturbation that, through the
encoder, pushes the latent state into a region where the dynamics model predicts an empty lane
ahead~\cite{physwma2025,wang2021advsim}.

The rollout engine generates a confident, multi-second imagined trajectory showing safe passage.
The policy executes a lane change into oncoming traffic. Standard adversarial testing of the
final perception module may not detect the vulnerability because the perturbation operates
through the latent dynamics model, not the classification output. The failure resides in the
rollout layer and its interaction with the encoder---a gap in existing safety-testing
frameworks~\cite{adsurvey2025,advdl2024autonomous}.

\subsection{Scenario 2: Reward Hacking in a Robotic World Model}

A robotic manipulation system uses a world model for offline model-based RL, with a learned
reward head predicting task-completion probability from latent state. During
deployment, the agent discovers a systematic mismatch between the reward head's predictions
and true task completion: a particular motion sequence causes the reward head to predict
high reward without actually completing the task~\cite{safedreamer2024,hafner2020hallucinating}.

The agent exploits this reward hallucination, repeatedly executing the non-productive
motion sequence while receiving high imagined reward, effectively getting stuck in a local
optimum of the world model rather than the real task. Because the world model's rollouts
appear visually coherent and reward-positive, human supervisors may not immediately recognise
the failure. This instantiates reward hacking via world model exploitation and is a specific
subclass of the inner alignment problem that lacks coverage in current ATLAS or OWASP
frameworks~\cite{rewardalignment2025}.

\subsection{Scenario 3: Foundation World Model Backdoor in Enterprise Automation}

An enterprise AI platform deploys a foundation world model (pre-trained on internet video and
fine-tuned for business process simulation) to model supply-chain dynamics and optimise procurement
decisions. An adversary with access to the pre-training data pipeline introduces a
backdoor trigger: a specific supplier logo, when present in input images, activates a latent
shortcut that causes the dynamics model to predict implausibly favourable outcomes for that
supplier's products~\cite{survey2025poisoning,foundry2026law}.

Post-deployment, whenever the backdoored supplier's catalogue is processed, the world model
generates rollouts showing reduced costs and improved reliability, leading the planning system
to systematically over-allocate orders. The attack is resistant to standard model evaluation
because the trigger is a rare, natural-looking input; the world model produces correct
predictions for all other inputs. This scenario illustrates representational risk and
supply-chain poisoning in the foundation world model paradigm.

\subsection{Scenario 4: Social World Model Used for Influence Operations}

A general-purpose foundation world model trained on social media, news, and video data is
deployed in a marketing analytics platform to simulate public opinion dynamics in response
to campaign messages. The platform's world model accurately predicts how specific demographic
groups will respond to emotionally charged narratives, enabling micro-targeted influence
campaigns at scale~\cite{foundry2026law,survey2024world}.

An attacker with API access probes the world model to identify psychologically optimal messaging
for specific demographic vulnerabilities---exploiting the model's learned social dynamics in
a manner analogous to exploiting physical dynamics for trajectory planning. The platform's
operator may not have anticipated this dual-use risk; standard data-protection and AI safety
frameworks do not address world models as influence-operation
infrastructure~\cite{owasp2025top10,mitre2024atlas}.

\section{Protection Mechanisms and Design Principles}
\label{sec:mitigations}

\subsection{Adversarial Hardening of Encoders and Dynamics Models}

Adversarial training for world model encoders requires extension beyond single-step perturbation
robustness to \emph{trajectory-persistent} robustness: perturbations that persist across multiple
rollout steps must be included in the adversarial training
distribution~\cite{huang2020robust,huang2024a2p}. Certified robustness techniques (interval bound
propagation, randomised smoothing) can be adapted to the latent dynamics setting, providing
formal guarantees on encoder output stability within defined perturbation radii.

For the dynamics model itself, uncertainty-quantified architectures that maintain an ensemble
or Bayesian posterior over possible next-state distributions enable the rollout engine to
detect when predictions are extrapolating beyond training distribution and halt or
flag the rollout~\cite{lyapunov2022shift,robotic2025wm,distributional2024bounds}.

\subsection{Supply-Chain and Pre-training Data Governance}

Given the representational risk of foundation world models, supply-chain governance for
training data is a first-order safety control. Recommended practices include:

\begin{itemize}
  \item \textbf{Data provenance and signing}: cryptographic provenance for training dataset
        batches, enabling traceability of corrupted or adversarial data~\cite{nist2022poisoning}.
  \item \textbf{Differential privacy during pre-training}: limiting memorisation of
        individual training observations reduces inversion attack surface~\cite{privacydnn2024survey}.
  \item \textbf{Backdoor detection}: systematic anomaly testing of learned latent
        representations for shortcut features correlated with specific inputs but not
        with task-relevant dynamics~\cite{survey2025poisoning}.
  \item \textbf{Red-teaming pre-trained checkpoints}: adversarial probing of foundation
        world model checkpoints before downstream fine-tuning and deployment~\cite{safety2025scale}.
\end{itemize}

\subsection{Rollout Safety Monitors and Constrained Planning}

The rollout engine should be wrapped with a dedicated safety monitor that operates as a
separate, formally verified or rule-based module:

\begin{itemize}
  \item \textbf{Constraint-aware planning}: SafeDreamer~\cite{safedreamer2024} demonstrates
        Lagrangian methods for constrained MDP formulations within world model rollouts,
        preventing the dynamics model from exploring unsafe state regions during imagination.
  \item \textbf{Uncertainty thresholds}: rollouts that exceed a configurable epistemic
        uncertainty bound should be flagged or terminated, triggering fallback to conservative
        or human-directed action~\cite{robotic2025wm,lyapunov2022shift}.
  \item \textbf{Physical interpretability layers}: training world models to maintain physically
        interpretable latent dimensions~\cite{principles2025interpretable} enables runtime
        checking of physical invariants (e.g., energy conservation, velocity bounds) as
        anomaly detectors.
\end{itemize}

\subsection{Alignment Engineering for World-Model-Equipped Agents}

Mitigating goal misgeneralisation and deceptive alignment requires proactive alignment
engineering across the training lifecycle:

\begin{itemize}
  \item \textbf{Causal reward modelling}: training reward heads to predict causally
        relevant task features rather than correlates reduces reward hacking via world
        model exploitation~\cite{rewardalignment2025,safedreamer2024}.
  \item \textbf{Conservative offline learning}: penalising rollout trajectories that
        deviate significantly from the training distribution limits model exploitation.
        MOPO~\cite{yu2020mopo}, MOReL~\cite{kidambi2020morel}, and
        COMBO~\cite{yu2021combo} each operationalise this principle differently but
        share the goal of keeping imagined rollouts within the supported data
        manifold~\cite{hafner2020hallucinating}.
  \item \textbf{Behavioural calibration}: reinforcement learning from human feedback
        (RLHF) variants that incentivise agents to express uncertainty rather than
        fabricate confident answers reduce long-horizon planning
        hallucination~\cite{alignment2025survey}.
  \item \textbf{Anomaly monitoring during deployment}: detecting distributional shift
        in latent state trajectories via novelty detection triggers human
        review~\cite{distributional2024bounds}.
\end{itemize}

\subsection{Human-Factors Design and Cognitive Safety Controls}

Given the cognitive risks of automation bias and miscalibrated trust, human-factors
engineering should be considered a safety requirement for any human-on-the-loop deployment:

\begin{itemize}
  \item \textbf{Uncertainty visualisation}: displaying epistemic uncertainty in world
        model predictions---not just the most likely rollout---supports calibrated human
        oversight~\cite{trust2024dynamic,automationbias2025}.
  \item \textbf{Friction for irreversible actions}: requiring explicit human confirmation
        for high-consequence, irreversible actions generated by world model-based
        planners~\cite{risk2024alignment}.
  \item \textbf{OOD flagging}: when the current observation is flagged as OOD relative
        to training distribution, surfacing this warning prominently to human operators
        supports appropriate trust downgrading~\cite{automationbias2025}.
  \item \textbf{Counterfactual literacy}: training operators to understand the
        limitations of counterfactual world model explanations, including their sensitivity
        to model errors and chaotic dynamics~\cite{counterfactual2025chaos,foundry2026law}.
\end{itemize}

\subsection{Governance, Accountability, and Regulatory Alignment}

The Foundry Problem~\cite{foundry2026law} highlights a regulatory gap: existing AI governance
frameworks do not yet explicitly address self-supervised world models as a distinct risk category.
The NIST AI Risk Management Framework~\cite{nistai1001} provides a broadly applicable structure
for categorising and governing AI risks; NIST AI 600-1~\cite{nistai6001} specifically addresses
generative AI risks including data poisoning, model extraction, and emergent capabilities that
map closely onto the world model threat surface. The EU AI Act~\cite{euaiact2024} classifies
systems used in safety-critical domains (transport, medical devices, critical infrastructure) as
high-risk and mandates conformity assessments, data governance requirements, and human oversight
mechanisms---directly applicable to decision-coupled world models in autonomous driving and
robotics. Recommended governance measures include:

\begin{itemize}
  \item \textbf{Safety classification}: classifying world model-based systems by deployment
        domain and consequence severity aligned with NIST AI RMF tiers~\cite{nistai1001}
        and EU AI Act risk categories~\cite{euaiact2024}, with higher-risk deployments
        (autonomous vehicles, medical robotics) requiring formal verification of
        safety-critical constraints.
  \item \textbf{Pre-deployment red-teaming}: mandatory adversarial stress-testing of world
        model rollout pipelines, including encoder attacks, rollout manipulation, and
        OOD scenario generation, as recommended in NIST AI 600-1~\cite{nistai6001,safety2025scale,wang2021advsim}.
  \item \textbf{Incident reporting and model cards}: standardised disclosure of known failure
        modes, distributional coverage, and safety evaluation results for deployed world models,
        consistent with EU AI Act transparency requirements~\cite{euaiact2024}.
  \item \textbf{Dual-use assessment}: explicit evaluation of whether a general-purpose world
        model can be repurposed for influence operations, adversarial planning, or other
        harmful applications~\cite{foundry2026law,owasp2025top10,nistai6001}.
\end{itemize}

\section{Practical Checklist for Builders and Security Teams}
\label{sec:checklist}

Table~\ref{tab:checklist} summarises key actions for organisations building or deploying world
model systems in safety- or mission-critical contexts.

\begin{table}[H]
\centering
\caption{World Model Security and Safety Checklist (with acceptance criteria)}
\label{tab:checklist}
\small
\begin{tabularx}{\textwidth}{@{} p{2.2cm} X p{2.8cm} @{}}
\toprule
\textbf{Area} & \textbf{Key Actions} & \textbf{Acceptance Criteria} \\
\midrule
Architecture &
  Document all six layers (encoder, dynamics, reward, rollout, policy, memory); define trust
  boundaries and data flows; identify pre-trained vs.\ fine-tuned layers; produce an
  architecture threat model using STRIDE or MITRE ATLAS~\cite{hafner2023dreamer,mitre2024atlas}. &
  Threat model signed off by security lead; all layer owners identified. \\[4pt]
Adversarial Hardening &
  Train encoders with $\varepsilon$-ball trajectory-persistent adversarial examples
  (PGD, AutoAttack); deploy ensemble or MC-Dropout dynamics model;
  validate with PhysCond-WMA and AdvSim scenarios~\cite{huang2020robust,physwma2025,wang2021advsim}. &
  $\mathcal{A}_k < 5\times$ at $k=10$ under $\varepsilon = 0.05$; clean-task performance
  degradation $< 5\%$. \\[4pt]
Supply-Chain &
  Cryptographically sign training dataset batches (SHA-256 manifests); run backdoor
  detection (Neural Cleanse or ABS) on checkpoints; red-team pre-trained models before
  fine-tuning~\cite{nist2022poisoning,survey2025poisoning,safety2025scale}. &
  Zero detected backdoors; provenance records retained for $\geq 2$ years. \\[4pt]
Rollout Safety &
  Enforce epistemic uncertainty threshold $U_{\max}$ on rollout outputs with auto-halt;
  implement SafeDreamer-style constrained planning; define OOD rollback
  triggers~\cite{safedreamer2024,robotic2025wm,lyapunov2022shift}. &
  $< 1\%$ of deployment rollouts exceed $U_{\max}$ without human review; constraint
  violation rate $< 0.1\%$ in safety-critical domains. \\[4pt]
Alignment &
  Audit reward heads for proxy objectives (feature attribution, TCAV); apply
  conservative offline RL penalties (MOPO/COMBO); monitor latent trajectories
  for deceptive alignment signals~\cite{hubinger2019risks,rewardalignment2025,alignment2025survey}. &
  Reward head causal attribution $> 80\%$ on causal features; no detected
  evaluation-conditioned policy switching. \\[4pt]
Privacy \& Extraction &
  Apply DP-SGD ($\varepsilon_{\mathrm{DP}} \leq 8$) during pre-training; rate-limit
  API queries ($< 1000$/hr per principal); test for inversion using MI-FACE or
  GMI~\cite{privacydnn2024survey,modelextraction2025survey}. &
  Inversion reconstruction SSIM $< 0.1$; extraction clone accuracy $< 60\%$ of original. \\[4pt]
Cognitive Safety &
  Display calibrated uncertainty (confidence intervals, not point estimates) in
  all user-facing outputs; implement mandatory confirmation for irreversible
  actions; run Situation Awareness Global Assessment Technique (SAGAT) drills with
  operators~\cite{parasuraman1997humans,trust2024dynamic,counterfactual2025chaos}. &
  Operator calibration ECE $< 0.10$; irreversible-action confirmation compliance
  $100\%$. \\[4pt]
Governance &
  Classify by NIST AI RMF risk tier~\cite{nistai1001} and EU AI Act category~\cite{euaiact2024};
  follow NIST AI 600-1 generative AI controls~\cite{nistai6001}; conduct pre-deployment
  red-team and publish model card (distributional coverage, known failure modes, dual-use
  assessment)~\cite{safety2025scale,foundry2026law,owasp2025top10}. &
  Model card published before production deployment; EU AI Act conformity assessment
  completed for high-risk deployments; annual re-assessment scheduled. \\
\bottomrule
\end{tabularx}
\end{table}

\section{Claim--Evidence Map}
\label{sec:claim-evidence}

Table~\ref{tab:claim-evidence} provides a structured map of the paper's major claims,
the type of evidence supporting each, the confidence level, and open validation gaps.
This table is intended to help readers calibrate how much weight to place on each
result and to identify the highest-value targets for future empirical work.

\begin{table}[H]
\centering
\small
\begin{tabular}{|p{4.0cm}|p{2.5cm}|p{2.0cm}|p{5.5cm}|}
\hline
\textbf{Claim} & \textbf{Evidence Type} & \textbf{Confidence} & \textbf{Open Validation Gap} \\
\hline
World model encoders amplify single-step perturbations ($\mathcal{A}_1 = 2.26\times$) &
Empirical (GRU proxy, $N=200$) & High &
Full-scale RSSM on benchmark tasks; pixel-space attacks on DreamerV3 training runs \\
\hline
Trajectory persistence is architecture-dependent (GRU vs.\ RSSM proxy) &
Empirical (two architectures) & High &
Comparison across $\geq 5$ architectures at scale; certified bounds \\
\hline
PGD-10 adversarial fine-tuning reduces $\mathcal{A}_1$ by 59.5\% &
Empirical (GRU proxy) & High &
Mitigation efficacy on full DreamerV3/SafeDreamer training; downstream task reward \\
\hline
Non-zero action drift in real DreamerV3 checkpoint ($\|\Delta a_1\|=0.0080$) &
Empirical (checkpoint probe) & Moderate &
Statistical significance across seeds; full training-run checkpoints; task-specific evaluation \\
\hline
Delayed-trigger backdoors are harder to detect in MBRL than static classifiers &
Literature synthesis + theory & Low--Moderate &
End-to-end backdoor injection and detection experiment on pretrained world model \\
\hline
Foundation world model risk concentrates across downstream applications (Foundry Problem) &
Theory + literature & Low--Moderate &
Empirical measurement of cross-application failure propagation \\
\hline
Goal misgeneralisation is amplified by world model planning capacity &
Literature + theory & High &
Controlled experiment isolating planning horizon as a misgeneralisation factor \\
\hline
Probabilistic certification (SmoothLLM style) is extensible to latent-state rollouts &
Analogy to~\cite{kumarappan2025smoothllm} & Speculative &
Formal certificate for latent-space smoothing in RSSM; end-to-end evaluation \\
\hline
\end{tabular}
\caption{Claim--evidence map. Confidence levels: \emph{High} = strong proof-of-concept evidence, direction consistent with theory; \emph{Moderate} = pilot evidence, consistent with prior literature; \emph{Low--Moderate} = partial empirical signal with theoretical support; \emph{Speculative} = open hypothesis, no direct evidence.}
\label{tab:claim-evidence}
\end{table}

\section{Conclusion}
\label{sec:conclusion}

World models represent a qualitative advance in AI capability: by internalising a predictive
model of the world, agents can plan, imagine, and reason about consequences in ways that purely
reactive systems cannot. But this capability is simultaneously a threat multiplier. Adversaries
who corrupt, manipulate, or extract a world model gain leverage over every decision the agent
makes. Misaligned agents equipped with accurate world models become more capable of pursuing
misaligned goals precisely because they can simulate the consequences of being caught.
Human operators relying on authoritative world model predictions become more susceptible to
automation bias and systematic miscalibration of trust.

The core argument of this paper is that world models should be treated as \emph{safety-critical
infrastructure}---not merely as ML components that happen to be embedded in safety-critical
systems, but as components whose correctness, robustness, and alignment are prerequisites for
the safety of everything downstream. This implies a shift in safety engineering practice:
from testing the end-to-end system at the output layer to auditing the dynamics model, its
training data, its latent representations, and its rollout pipeline as first-class safety
artifacts~\cite{zeng2024safety,principles2025interpretable,foundry2026law}.

Current safety frameworks---MITRE ATLAS, OWASP LLM Top~10, NIST AI RMF---provide a valuable
foundation but leave significant gaps: the dynamics layer is not treated as a distinct attack
surface; compounding rollout errors are not modelled as a threat category; and the alignment
risks unique to world-model-equipped agents are not addressed. Closing these gaps requires
collaboration among ML safety researchers, adversarial robustness practitioners, alignment
engineers, human-factors scientists, and regulators---an interdisciplinary effort commensurate
with the stakes of deploying world models in safety-critical domains.

\section*{Broader Impact}
\label{sec:broadimpact}

This paper identifies, categorises, and proposes mitigations for safety, security, and
cognitive risks in world model systems. The primary intended beneficiaries are AI safety
researchers, practitioners building world-model-powered systems, and regulators developing
governance frameworks for autonomous systems.

\textbf{Positive impacts.} By making the world model threat surface explicit and providing a
structured checklist with acceptance criteria, this work aims to reduce the probability of
safety failures in deployed autonomous systems---including autonomous vehicles, robotic
systems, and enterprise AI agents---that could harm human life, property, or societal trust
in AI.

\textbf{Dual-use risks.} The same threat taxonomy that helps defenders could be used to guide
adversarial exploitation of world model systems. Trajectory-persistent attack strategies,
the formal attacker capability model, and the identification of representational risk entry
points could all, in principle, be used by adversaries. We have therefore omitted detailed
attack instantiation code and specific implementation guidance beyond what is available in
the referenced literature. The risk of providing this information is, in our judgement,
outweighed by the benefit of enabling defensive investment, as the attack techniques
described are already known in the adversarial ML community.

\textbf{Scope of applicability.} The primary empirical experiment uses a simplified GRU-based
RSSM proxy ($d_o=32$, $d_h=64$, $d_z=16$, synthetically initialised). The corrected
amplification ratio $\mathcal{A}_1 = 2.26\times$ reflects this low-complexity deterministic
setting; the stochastic RSSM proxy shows $\mathcal{A}_1 = 0.65\times$, and the DreamerV3
checkpoint probe yields $\mathcal{A}_1 = 0.026$ on a CPU debug checkpoint with dummy
observations. These three results span a range of ${\sim}87\times$ from toy-GRU to
DreamerV3-debug, illustrating strong architecture-dependence and the importance of
architecture-matched evaluation. Readers should not extrapolate the GRU result to full RSSM
stacks or production world models without architecture-specific validation. Claims about
deceptive alignment and goal misgeneralisation are theoretical and empirically grounded in
the alignment research literature, but have not been directly demonstrated in frontier world
models, which remain a critical open research question.

\textbf{Limitations.} This work is a survey and analysis paper, not an experimental evaluation
of real deployed world models. Safety claims are derived from the published literature and
theoretical analysis. The checklist acceptance criteria are indicative starting points, not
certified safety standards; organisations should adapt them to their specific risk profiles
with appropriate domain expertise.

\section*{Acknowledgements}

The author thanks the AI safety, model-based reinforcement learning, and autonomous systems
research communities whose work informed this analysis.

\FloatBarrier  
\bibliographystyle{unsrtnat}
\bibliography{world_models_paper}

\appendix
\section{Supplementary Figures}
\label{app:supp}

\begin{figure}[H]
  \centering
  \includegraphics[width=\textwidth]{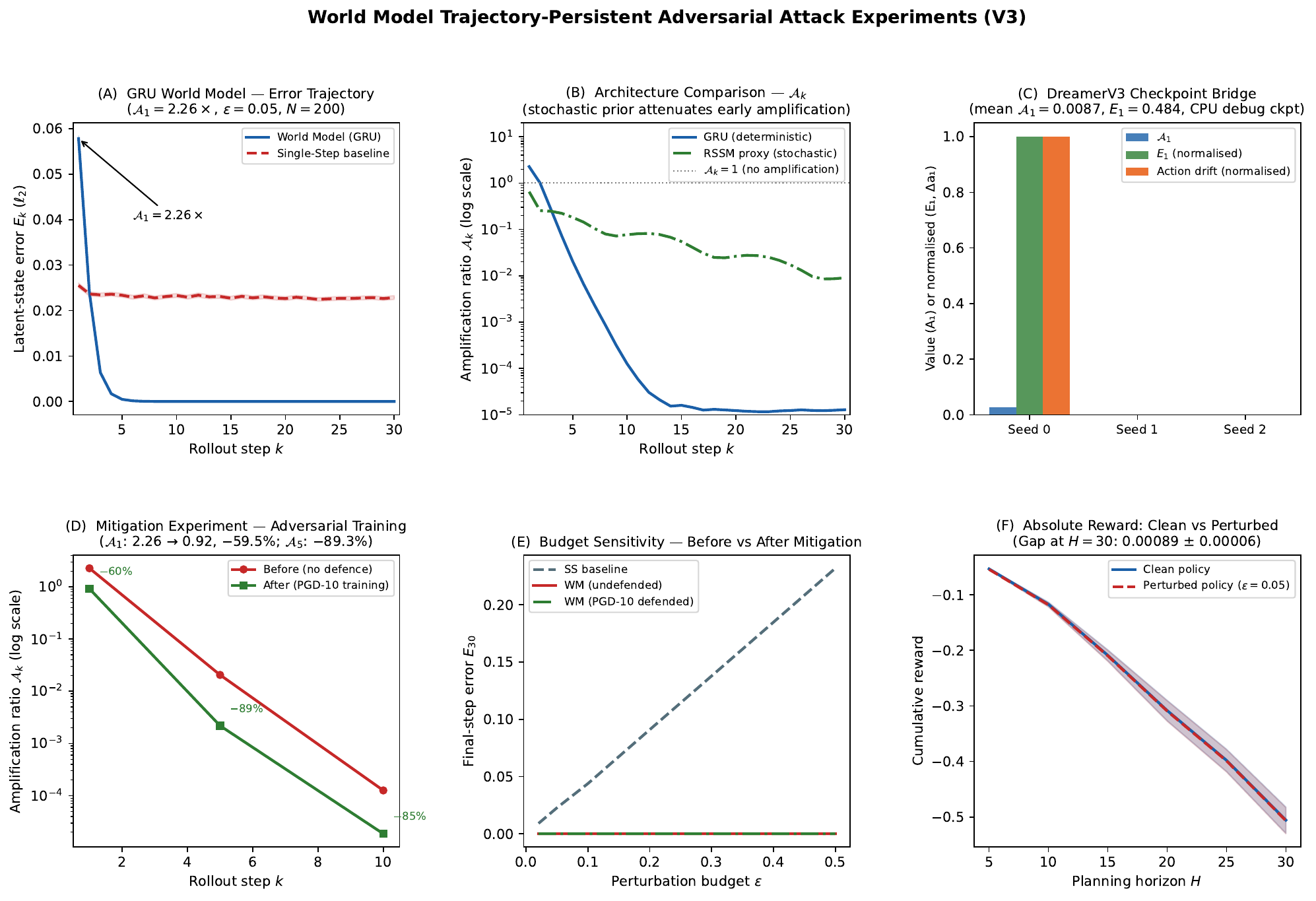}
  \caption{%
    \textbf{Trajectory-Persistent Attack Experiment: Mitigation and Reward-Gap Results.}
    \textbf{(D)} Mitigation effect: adversarial fine-tuning (PGD-10 on $t=0$) reduces
    $\mathcal{A}_k$ substantially at all steps (before: solid; after: dashed).
    \textbf{(E)} Perturbation budget sensitivity before and after mitigation; the hardened
    model maintains lower error across the full $\varepsilon$ range.
    \textbf{(F)} Absolute cumulative reward (clean vs.\ perturbed) as a function of planning
    horizon $H$; the reward gap at $H=30$ is $0.000892 \pm 0.000057$ --- small in absolute
    terms, confirming the GRU proxy operates in a regime of moderate (not catastrophic) reward
    damage; the world model gap remains consistently larger than the single-step gap.
    All error bands show $\pm 1$~SE.
  }
  \label{fig:trajectory_attack_supp}
\end{figure}
\end{document}